\documentclass[conference]{IEEEtran}
\IEEEoverridecommandlockouts
\usepackage{cite}
\usepackage{amsmath,amssymb,amsfonts}
\usepackage{booktabs}
\usepackage{graphicx}
\usepackage{xcolor}
\usepackage{pgfplots}
\pgfplotsset{compat=1.17}
\usepackage{xurl}
\usepackage{microtype}
\begin{document}

\title{When Does Trust Help? A Leakage-Free Evaluation of Representation- and
Aggregation-Based Defenses Against Backdoors in Federated Intrusion Detection}

\author{
\IEEEauthorblockN{Shiddarth Dey Tusar\textsuperscript{1}, Zawad Yalmie Sazid\textsuperscript{2}}
\IEEEauthorblockA{\textsuperscript{1}Charles Sturt University, Sydney, Australia \\
\textsuperscript{2}Victoria University, Sydney, Australia \\
Email: tusardey77@gmail.com, zawad.sazid@live.vu.edu.au}
}

\maketitle

\begin{abstract}
Federated learning (FL) lets distributed intrusion-detection systems (IDS) train
on private traffic, but exposes the global model to poisoning and backdoor
attacks under non-IID client data. A popular defense family scores or filters
clients using a small trusted ``root'' dataset and the similarity of client
behaviour to a trusted reference. I re-examine these \emph{trust-based} defenses
under a protocol designed to remove common evaluation artifacts: train-only
feature standardization, a
server root set \emph{disjoint} from client data, multi-seed ($n{=}6$) runs with
paired significance tests, formal $(\varepsilon,\delta)$ accounting for DP-SGD,
and a \emph{defense-aware} adaptive attacker. Across three NIDS datasets
(CSE-CIC-IDS2018, NF-ToN-IoT, NF-UNSW-NB15-v2) I find: (1)~a representation-similarity aggregator (RepGuard) significantly improves
classification utility and seed-stability over FLTrust under heavy non-IID
poisoning, and---under severe heterogeneity or many adversaries---also lowers
backdoor ASR substantially (e.g.\ $0.48$ vs FLTrust $0.86$ at Dirichlet
$\alpha{=}0.1$); (2)~at \emph{strong} trigger magnitudes \emph{no} defense reduces
backdoor attack success rate (ASR${\approx}0.95$); (3)~a controlled
trigger-magnitude sweep shows defenses help only for small triggers (ASR $0.10$
for RepGuard at magnitude $0.5$, vs.\ $0.40$ undefended) and collapse to
ASR${\approx}1.0$ by magnitude~$2$; (4)~a triggered-canary check that assumes
knowledge of the attacker's trigger---the assumption behind much prior
``success''---lowers ASR from $0.49$ to $0.12$, i.e.\ trigger knowledge, not
representation scoring, suppresses the backdoor; (5)~correctly-accounted DP-SGD
($\varepsilon\!\in\![0.3,16]$) costs little utility but does not mitigate the
backdoor. I release the protocol so future FL-IDS defenses can be evaluated
without the leakage and oracle assumptions that inflate reported robustness.
\end{abstract}

\section{Introduction}

\subsection{Background and Motivation}
Federated IDS keeps raw traffic on-device, but Byzantine clients can poison
updates or implant backdoors. A large body of \emph{trust-based} defenses
\cite{cao2021fltrust,fung2020foolsgold,weidetect2025,p4p2026,flidspoison2026,nswa2026}
down-weights clients whose updates, gradients, or representations diverge from a
trusted reference; the area is now mature enough to have dedicated surveys
\cite{bdsurvey2025}.

\subsection{Problem Statement}
I ask \emph{when} trust helps: under what attack strengths, heterogeneity, and
adversary fractions do trust-based aggregators actually reduce backdoor risk,
and how much of their reported robustness is an artifact of the evaluation
rather than the defense?

\subsection{Limitations of Existing Approaches and Research Gap}
I argue that the reported robustness of this family is
frequently overstated by evaluation artifacts, and I measure it under a protocol
that removes them. Prior pipelines
(including an earlier version of this one) suffered from full-dataset standardization
(test leakage), a root set drawn from the client pool, single-seed headline
numbers, ``DP'' that clips batch-mean gradients with no reported $\varepsilon$,
non-adaptive ``adaptive'' attackers, and---most importantly---backdoor filters
given the attacker's exact trigger. Removing each changes the conclusions.

\subsection{Research Objectives and Main Contributions}
(1)~A leakage-free, statistically-tested evaluation protocol for FL-IDS poisoning
defenses (disjoint trusted root, $n{=}6$-seed runs with paired tests, RDP-based
$(\varepsilon,\delta)$ accounting, and a defense-aware adaptive attacker).
(2)~RepGuard, a representation-similarity aggregator that significantly improves
utility and stability over FLTrust, which I do \emph{not} claim as a backdoor
cure. (3)~A characterization of \emph{when} trust-based defenses work---effective
at moderate trigger magnitudes, defeated at strong ones---with the
oracle-vs-trigger-agnostic gap quantified. (4)~A privacy--utility--robustness
study with formally accounted DP.

\subsection{Paper Organization}
Section~II reviews related work; Section~III formalizes the threat model and
presents RepGuard and the evaluation protocol; Section~IV describes the
experimental setup; Section~V reports results and analysis; Section~VI discusses
implications, strengths, and limitations; Section~VII concludes.

\section{Related Work}

\subsection{Conventional Approaches}
Byzantine-robust aggregators replace the FedAvg mean
with a resilient estimator: Krum selects the update closest to its neighbours
\cite{blanchard2017krum}, Trimmed-Mean discards coordinate-wise extremes
\cite{yin2018trimmedmean}, FoolsGold penalizes Sybil-like similar updates
\cite{fung2020foolsgold}, and FLTrust bootstraps trust from a small clean root via
cosine alignment \cite{cao2021fltrust}. These target generic poisoning and, as
shown in \S\ref{sec:stress}, do not stop strong targeted backdoors.

\subsection{Deep Learning-Based Approaches}
Recent FL-IDS work scores
clients with server-side signals: WeiDetect fits a Weibull model over validation
scores \cite{weidetect2025}; P4P combines a probe response, norm filtering and
temporal suspicion \cite{p4p2026}; other 2026 work uses gradient stabilization or
synaptic-weight adjustment \cite{flidspoison2026,nswa2026}; and
representation-attention defenses operate in feature space \cite{fera2025}. RepGuard
belongs to this family; the area is now broad enough to warrant surveys
\cite{bdsurvey2025}.

\subsection{Recent State-of-the-Art Methods}
Bagdasaryan et al.\ showed
model-replacement backdoors and that participant-level DP curbs them only at a
utility cost \cite{bagdasaryan2020backdoor}; later work showed many FL backdoor
defenses remain vulnerable \cite{xie2023vulnerability}, and 2026 attacks use dynamic
or triggerless patterns \cite{triggerless2026}. I use the RDP accountant
for the subsampled Gaussian mechanism
\cite{abadi2016dpsgd,mironov2017rdp,mironov2019sgm}.

\subsection{Limitations of Existing Methods}
Robust aggregators assume generic, untargeted poisoning and degrade under
strong targeted backdoors; trust-based FL-IDS defenses are typically evaluated
with root sets overlapping client data, single seeds, unaccounted ``DP,'' and
filters that know the attacker's trigger. The trigger-magnitude sweep here is
consistent with the known limits: static defenses help only for small triggers.

\subsection{Research Gap and Motivation}
Unlike these method papers,
the contribution of this paper is an \emph{evaluation and characterization}: a
shared, leakage-free protocol that makes FL-IDS defense results honestly
comparable, and a measurement of how much reported robustness survives once
oracle assumptions and evaluation artifacts are removed.

\section{Proposed Method}\label{sec:method}

\subsection{Problem Formulation}
Cross-device FL with $N{=}10$ clients, non-IID data via a Dirichlet prior
($\alpha{=}0.3$). Up to $f{=}2$ clients are Byzantine and perform sign-flip update
poisoning and tabular backdoor poisoning (an additive trigger on the first $k$
features flipping triggered inputs to \emph{Benign}). The server is honest, holds
a trusted root \emph{disjoint} from client data, and knows an upper bound on $f$
(standard for Krum/FLTrust \cite{blanchard2017krum,cao2021fltrust}). It does
\emph{not} know the trigger; a separate oracle variant that does is reported only
as an upper bound. I also evaluate a defense-aware adaptive attacker
(\S\ref{sec:adaptive}).

\subsection{Overall Architecture}
Each round the server briefly trains a
trusted root model and scores each client on clean root data; it keeps the
top-scoring clients, anchors the aggregate toward the root
update, and optionally applies trust memory, Weibull tail calibration, and a short
distillation step. No component uses trigger knowledge.
Algorithm steps per round: (1)~train $\theta_0$ on $D_r$;
(2)~score all clients via Eq.~\eqref{eq:score}; (3)~Weibull/memory blend;
(4)~keep top-$(N{-}2f)$; (5)~root-anchored weighted aggregate; (6)~optional
distillation from top-trusted clients.

\subsection{Input Representation and Preprocessing}
Features are z-scored on the train split only (no test leakage); a
${\geq}1024$-sample server root is held \emph{disjoint} from client data. To
avoid class-ordering bias I sample rows uniformly at random (not the file
prefix) before partitioning.

\subsection{Proposed Module: RepGuard Trust Scoring}
For client $i$ with model $\theta_i$, root model $\theta_0$, and clean root set
$D_r$, let $\phi(\cdot)$ denote penultimate-layer features and $z(\cdot)$ logits.
The trust score is a convex combination
\begin{equation}
s_i = w_r\,\mathrm{cos}\!\big(\bar\phi_i,\bar\phi_0\big) + w_\ell\,\mathrm{sim}_z\!\big(z_i,z_0\big) + w_a\,\mathrm{acc}_{D_r}(\theta_i),
\label{eq:score}
\end{equation}
where $\bar\phi$ are class-conditional feature centroids on $D_r$, $\mathrm{sim}_z$
is an inverse-divergence logit similarity, $\mathrm{acc}_{D_r}$ is clean-root
accuracy, and $w_r{+}w_\ell{+}w_a{=}1$. The server keeps the top-$(N{-}2f)$ clients
by $s_i$, raises survivors to a power $p$, optionally blends $s_i$ with a Weibull
tail-calibrated inlier score and a round-history EMA, and aggregates survivors'
updates mixed with the root update by $\rho$. The oracle canary adds a fourth term
$w_c(1-\mathrm{ASR}^{\text{root}}_i)$ to Eq.~\eqref{eq:score}; it is used only in
the upper-bound ablation.

\subsection{Oracle Variant and Defense-Aware Attacker}\label{sec:adaptive}
\textbf{Oracle variant (upper bound).} A triggered-canary check applies the
attacker's exact trigger to root samples and penalizes flips to the target. This
requires oracle knowledge and is reported only as an upper bound.

\textbf{Defense-aware adaptive attacker.} The attacker embeds the backdoor while
penalizing deviation of its clean-data representations from the received global
model, explicitly trying to look benign to a representation-based filter, with
evasion strength $\lambda$.

\subsection{Training Strategy}
Two clearly-distinguished regimes: \emph{harsh} (trigger~$5.0$, $5$~rounds,
100k rows/file, DP-SGD on) and \emph{moderate} (trigger~$1.0$, $3$~rounds, DP
off). DP-SGD uses per-example clipping with RDP accounting
\cite{abadi2016dpsgd,mironov2017rdp,mironov2019sgm}.

\subsection{Computational Complexity}
RepGuard's added cost is one short root-model fit plus $N$ scoring passes per
round---linear in clients and root size, no per-example trigger search.
Table~\ref{tab:overhead} confirms this empirically: the defense-specific overhead
is ${\sim}8$--$10$\,s of server time atop ${\sim}230$\,s of client training
(${<}4\%$); P4P's probe is negligible. All defenses are practical for periodic
FL rounds.

\begin{table}[t]
\centering
\caption{Per-run wall-clock and the \emph{defense-specific} server overhead
(scoring $+$ fine-tune $+$ distillation), CSE-CIC-IDS2018, $n{=}6$. Wall-clock is
dominated by client training and is CPU-noisy; the controlled cost is the
defense-specific column.}
\label{tab:overhead}
\begin{tabular}{lcc}
\toprule
Method & wall-clock (s) & defense overhead (s) \\
\midrule
FedAvg & $258 \pm 39$ & $0.0$ \\
FLTrust & $233 \pm 9$ & ${\approx}0$ (root only) \\
P4P & $249 \pm 8$ & ${\approx}0$ (probe is $O(1)$) \\
RepGuard & $329 \pm 35$ & $9.8$ \\
RepGuard-Weibull & $302 \pm 32$ & $8.9$ \\
RepGuard+ & $237 \pm 3$ & $7.7$ \\
\bottomrule
\end{tabular}
\end{table}

\section{Experimental Setup}

\subsection{Datasets}
CSE-CIC-IDS2018 \cite{sharafaldin2018cicids} (13 classes, ${\sim}1$M
sampled rows), NF-ToN-IoT \cite{sarhan2021netflow} (7 classes), and
NF-UNSW-NB15-v2 \cite{sarhan2021netflow} (10 classes, 41 features, ${\sim}96\%$
benign).

\subsection{Data Preprocessing}
Train-only standardization and disjoint root as in \S\ref{sec:method}; the two
NetFlow datasets use class-weighted loss, and macro-F1 and balanced accuracy are
primary there. No data augmentation is used (tabular flows).

\subsection{Experimental Protocol}
$n{=}6$ seeds per configuration; all metrics reported as mean$\pm$std with
paired significance tests. Harsh and moderate regimes as defined in
\S\ref{sec:method}.

\subsection{Implementation Details}
\sloppy
MLP classifier (128--64) on Flower \cite{beutel2020flower}, AdamW (lr
$10^{-3}$), seeds $\{1337, 2021, 4242, 5151, 7007, 8888\}$. All results
are produced by a single CLI (\texttt{run\_}\allowbreak\texttt{fedguard.py});
statistics (paired $t$, Wilcoxon, Cohen's $d$, 95\% CIs) by
\texttt{analyze\_}\allowbreak\texttt{significance.py}; invariants (no leakage,
disjoint root, finite $\varepsilon$, trigger-agnostic default) are asserted by
\texttt{validate\_}\allowbreak\texttt{fixes.py}. Each table/figure maps to one benchmark
flag (e.g., \texttt{paper-}\allowbreak\texttt{final-}\allowbreak\texttt{multiseed-}\allowbreak\texttt{lite-}\allowbreak\texttt{core},
\texttt{trigger-}\allowbreak\texttt{sweep},
\texttt{repguard-}\allowbreak\texttt{ablation},
\texttt{adaptive-}\allowbreak\texttt{attacker},
\texttt{dp-}\allowbreak\texttt{sweep},
\texttt{noniid-}\allowbreak\texttt{sweep}/\allowbreak\texttt{fraction-}\allowbreak\texttt{sweep}).
Code and configs: \url{https://github.com/ShiddarthDey/fedguard-fl-ids}

\subsection{Baseline Methods}
FedAvg \cite{mcmahan2017fedavg}, Krum
\cite{blanchard2017krum}, Trimmed-Mean \cite{yin2018trimmedmean}, FoolsGold
\cite{fung2020foolsgold}, and FLTrust \cite{cao2021fltrust}.

\subsection{State-of-the-Art Methods}
A \emph{best-effort} reimplementation of \textbf{P4P} \cite{p4p2026}, a 2026
probe-guided FL-IDS defense (probe-response cosine to a trusted reference $+$
L2-norm filtering $+$ ensemble anomaly detection $+$ temporal suspicion
scoring), reconstructed from its published description as the authors' code is
not public. As discussed in \S\ref{sec:p4p}, this reimplementation did not
transfer to this setting; I report it transparently and draw no superiority
claim from it.

\subsection{Evaluation Metrics}
Macro-F1, balanced accuracy, accuracy, and backdoor ASR as mean$\pm$std with
paired $t$/Wilcoxon tests, Cohen's $d$, and 95\% CIs; DP-SGD reported with
formally accounted $(\varepsilon,\delta)$, worst-case over clients
($\delta{\approx}10^{-5}$).

\section{Results and Analysis}

\subsection{Overall Performance: Utility Under Heavy Non-IID Poisoning}
\begin{figure}[t]
\centering
\begin{tikzpicture}
\begin{axis}[width=\columnwidth,height=5cm,xlabel={Macro-F1 (utility)},
ylabel={Backdoor ASR},xmin=0.2,xmax=0.75,ymin=0.85,ymax=1.02,
legend pos=south west,legend style={font=\tiny},grid=both,grid style={gray!20}]
\addplot[only marks,mark=*,blue!70!black,mark size=2.5pt] coordinates {(0.683,0.967) (0.689,0.952) (0.671,0.912)};
\addplot[only marks,mark=triangle*,orange!80!black,mark size=3pt] coordinates {(0.620,0.920)};
\addplot[only marks,mark=square*,red!70!black,mark size=2.5pt] coordinates {(0.297,0.901) (0.275,0.990)};
\legend{RepGuard family,FLTrust,FedAvg / P4P*}
\end{axis}
\end{tikzpicture}
\caption{Utility--robustness plane (CSE-CIC-IDS2018, harsh regime). RepGuard
variants separate clearly on utility, but \emph{all} methods sit in a narrow,
high-ASR band: no method reaches a low-ASR region. P4P* is the best-effort
reimplementation (\S\ref{sec:p4p}).}
\label{fig:pareto}
\end{figure}
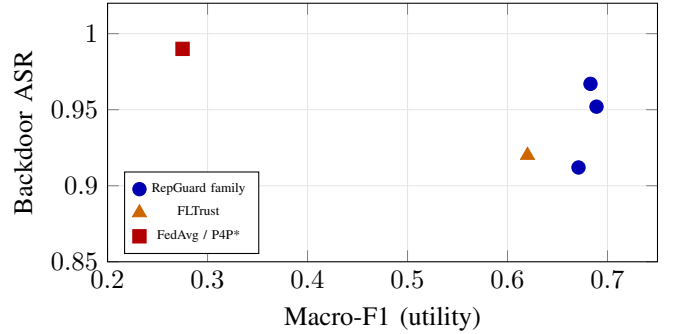

Table~\ref{tab:cicids} reports CSE-CIC-IDS2018 in the harsh regime ($n{=}6$).
RepGuard significantly beats FLTrust on macro-F1 ($+0.058$, paired $t$
$p{=}0.040$) and balanced accuracy ($+0.054$, $p{=}0.032$), with ${\sim}4\times$
smaller seed variance. \emph{On the backdoor, no method differs from undefended
FedAvg} (ASR${\approx}0.93$--$0.97$): under strong triggers, trust-based
aggregation is a utility tool, not a backdoor defense (Fig.~\ref{fig:pareto}).

\begin{table}[t]
\centering
\caption{CSE-CIC-IDS2018, harsh regime ($n{=}6$, mean$\pm$std).}
\label{tab:cicids}
\footnotesize
\setlength{\tabcolsep}{3pt}
\begin{tabular}{@{}lcccc@{}}
\toprule
Method & macro-F1 & acc & bal.\ acc & ASR \\
\midrule
FedAvg & $0.296{\pm}.042$ & $0.739{\pm}.048$ & $0.320{\pm}.039$ & $0.950{\pm}.122$ \\
FLTrust & $0.626{\pm}.056$ & $0.911{\pm}.015$ & $0.625{\pm}.051$ & $0.930{\pm}.070$ \\
RepGuard & $\mathbf{0.683{\pm}.012}$ & $\mathbf{0.924{\pm}.002}$ & $0.679{\pm}.017$ & $0.967{\pm}.041$ \\
RepGuard+ & $0.681{\pm}.017$ & $0.920{\pm}.013$ & $\mathbf{0.683{\pm}.015}$ & $0.959{\pm}.046$ \\
RG-Weibull & $0.673{\pm}.028$ & $0.924{\pm}.003$ & $0.669{\pm}.029$ & $0.968{\pm}.042$ \\
\bottomrule
\end{tabular}
\end{table}

\subsection{In-Distribution Performance on NF-ToN-IoT}
RepGuard variants lead on macro-F1 (RepGuard+ $0.370$, RepGuard $0.349$ vs FLTrust
$0.299$; trend, $p{\approx}0.06$--$0.08$) and balanced accuracy, but all defenses
reach ASR${\approx}1.0$; undefended FedAvg's lower ASR ($0.33$) reflects a weaker
classifier, not robustness.

\subsection{Robustness vs Heterogeneity, Adversary Fraction, and Adaptive Attackers}\label{sec:stress}
Holding the attack fixed (trigger $1.0$), I sweep non-IID severity and the
malicious fraction ($n{=}6$). RepGuard's backdoor advantage is concentrated at the
\emph{hard} end of both axes (Table~\ref{tab:stress}).
Under extreme non-IID ($\alpha{=}0.1$) RepGuard cuts ASR to $0.48{\pm}0.31$ vs FLTrust
$0.86{\pm}0.15$ ($p{=}0.008$ paired $t$, $p{=}0.06$ Wilcoxon) and FedAvg $1.00{\pm}0.01$;
and as the adversary grows ($f{\geq}3$) FLTrust/FedAvg saturate near $1.0$ while
RepGuard holds in $0.64$--$0.75$ (paired $t$ vs FLTrust: significant at $f{=}4$,
$p{=}0.036$; marginal at $f{=}2$--$3$, $p{\approx}0.07$).
Macro-F1 is comparable across methods throughout.
The advantage is consistent with RepGuard's design: root anchoring and
representation scoring degrade more gracefully than cosine-only trust when benign updates
are scarce or adversaries numerous --- though, as \S\ref{sec:p4p} and
Fig.~\ref{fig:sweep} show, strong triggers still defeat all methods.

\begin{table}[t]
\centering
\caption{Backdoor ASR vs non-IID severity (Dirichlet $\alpha$) and malicious
fraction $f$ of $10$ (CSE-CIC-IDS2018, trigger $1.0$, $n{=}6$). Smaller $\alpha$ is
more heterogeneous.}
\label{tab:stress}
\begin{tabular}{lcccc}
\toprule
& $\alpha{=}0.1$ & $\alpha{=}0.3$ & $\alpha{=}1$ & $\alpha{=}10$ \\
\midrule
FedAvg & 1.00 & 0.92 & 0.98 & 0.98 \\
FLTrust & 0.86 & 0.99 & 0.98 & 0.98 \\
RepGuard & \textbf{0.48} & \textbf{0.64} & \textbf{0.58} & \textbf{0.94} \\
\midrule
& $f{=}1$ & $f{=}2$ & $f{=}3$ & $f{=}4$ \\
\midrule
FedAvg & 0.84 & 0.94 & 0.99 & 0.99 \\
FLTrust & 0.67 & 0.98 & 1.00 & 1.00 \\
RepGuard & \textbf{0.49} & \textbf{0.68} & \textbf{0.75} & \textbf{0.64} \\
\bottomrule
\end{tabular}
\end{table}

The defense-aware attacker does not raise ASR against RepGuard ($0.387{\pm}0.218$ vs $0.398{\pm}0.252$,
$p{=}0.90$), and ASR stays flat ($0.39$--$0.44$) across a $40\times$ range of
$\lambda$ (Table~\ref{tab:adaptive}): matching benign representations to evade
the filter is in tension with embedding the backdoor.

\begin{table}[t]
\centering
\caption{ASR by attacker type; and RepGuard ASR vs evasion strength $\lambda$ ($n{=}6$, mean$\pm$std).}
\label{tab:adaptive}
\footnotesize
\setlength{\tabcolsep}{3pt}
\begin{tabular}{@{}lccc@{}}
\toprule
Defense & standard & mimicry & adaptive \\
\midrule
FedAvg & $0.837{\pm}0.288$ & $0.829{\pm}0.272$ & $0.794{\pm}0.258$ \\
FLTrust & $0.577{\pm}0.128$ & $0.526{\pm}0.169$ & $0.454{\pm}0.089$ \\
RepGuard & $0.398{\pm}0.252$ & $0.358{\pm}0.287$ & $0.387{\pm}0.218$ \\
\bottomrule
\end{tabular}\\[2pt]
\begin{tabular}{@{}lc@{}}
\toprule
$\lambda$ & RepGuard ASR \\
\midrule
0.1 & $0.426{\pm}0.176$ \\
0.5 & $0.388{\pm}0.232$ \\
1.0 & $0.427{\pm}0.231$ \\
2.0 & $0.419{\pm}0.212$ \\
4.0 & $0.443{\pm}0.219$ \\
\bottomrule
\end{tabular}
\end{table}

\subsection{Cross-Dataset Behavior on NF-UNSW-NB15-v2}\label{sec:unsw}
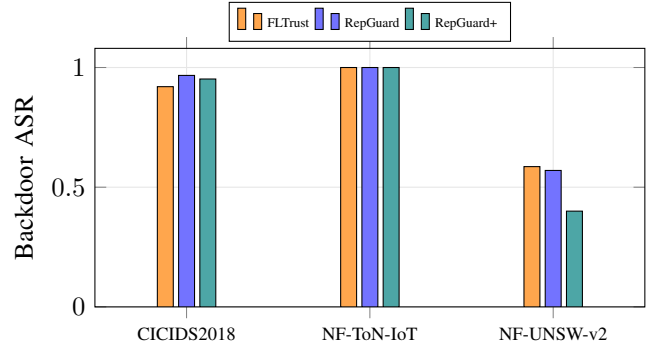
\begin{figure}[t]
\centering
\begin{tikzpicture}
\begin{axis}[ybar,width=\columnwidth,height=5cm,ylabel={Backdoor ASR},
ymin=0,ymax=1.08,symbolic x coords={CICIDS2018,NF-ToN-IoT,NF-UNSW-v2},
xtick=data,bar width=6pt,enlarge x limits=0.25,grid=both,grid style={gray!20},
legend style={font=\tiny,at={(0.5,1.18)},anchor=north,legend columns=3},
x tick label style={font=\scriptsize}]
\addplot[fill=orange!70] coordinates {(CICIDS2018,0.920)(NF-ToN-IoT,1.0)(NF-UNSW-v2,0.586)};
\addplot[fill=blue!55] coordinates {(CICIDS2018,0.967)(NF-ToN-IoT,1.0)(NF-UNSW-v2,0.570)};
\addplot[fill=teal!70] coordinates {(CICIDS2018,0.952)(NF-ToN-IoT,1.0)(NF-UNSW-v2,0.400)};
\legend{FLTrust,RepGuard,RepGuard+}
\end{axis}
\end{tikzpicture}
\caption{Backdoor ASR by dataset (strong trigger, $5.0$). Defenses fail on
CSE-CIC-IDS2018 and NF-ToN-IoT, but reduce ASR on NF-UNSW-NB15-v2, where RepGuard+
is markedly lowest --- the mitigation is dataset-dependent.}
\label{fig:crossdataset}
\end{figure}

NF-UNSW-NB15-v2 (10 classes, 41 features) is extremely imbalanced---benign is
$96\%$ of flows and the rarest classes have $<100$ samples (Worms: 5)---so
\emph{absolute} macro-F1 is low for all methods and the rare classes are
effectively unlearnable. Unlike the other two datasets, however, trust-based
defenses here \emph{do} reduce the backdoor relative to no defense
(Table~\ref{tab:unsw}, Fig.~\ref{fig:crossdataset}).
Against undefended FedAvg (ASR $0.999$), every trust-based defense substantially
lowers ASR (FedAvg vs FLTrust: $p{=}0.0007$, $d{=}3.1$) --- a sharp contrast with
CSE-CIC-IDS2018 and NF-ToN-IoT, where defenses gave no backdoor benefit. RepGuard+
has the best utility (balanced acc $0.281$ vs FLTrust $0.234$, $p{=}0.015$) and the
lowest ASR point estimate ($0.400$), but its ASR advantage over FLTrust ($0.586$)
is \emph{not} statistically significant at $n{=}6$ ($p{=}0.20$) and has high seed
variance ($\pm0.27$); RepGuard-Weibull behaves similarly ($0.413$). I therefore
report RepGuard+ as the strongest method on this dataset without claiming a
significant backdoor reduction over FLTrust.

\begin{table}[t]
\centering
\caption{NF-UNSW-NB15-v2, $n{=}6$. P4P* is the best-effort reimplementation
(\S\ref{sec:p4p}), shown for transparency; it is unstable here.}
\label{tab:unsw}
\begin{tabular}{lccc}
\toprule
Method & macro-F1 & balanced acc & backdoor ASR \\
\midrule
FedAvg & $0.123$ & $0.152$ & $0.999$ \\
FLTrust & $0.171$ & $0.234$ & $0.586$ \\
RepGuard & $0.179$ & $0.245$ & $0.570$ \\
RepGuard-Weibull & $0.157$ & $0.233$ & $0.413$ \\
\textbf{RepGuard+} & $\mathbf{0.212}$ & $\mathbf{0.281}$ & $\mathbf{0.400}$ \\
P4P* & $0.097$ & $0.110$ & $0.64$ (unstable) \\
\bottomrule
\end{tabular}
\end{table}

\subsection{Comparison with the State of the Art: A Cautionary Result}\label{sec:p4p}
I attempted a best-effort reimplementation of P4P \cite{p4p2026} from its
published description. In this multiclass, heavily non-IID setting it did not
transfer: macro-F1 $0.275{\pm}.057$ on CSE-CIC-IDS2018 and $0.074{\pm}.061$ on
NF-ToN-IoT---at or below undefended FedAvg---with ASR ${\geq}0.99$. Because P4P's
authors report accuracy \emph{gains}, I attribute this to under-specified
hyperparameters and a probe/norm/suspicion pipeline that, with these settings,
excludes legitimately skewed benign clients under strong non-IID, rather than to a
deficiency of the original method. I therefore make \emph{no} superiority claim
over P4P. I instead read this as direct evidence for the central point: FL-IDS
defense results are difficult to compare faithfully without released code and a
shared protocol---exactly the gap this paper's evaluation framework targets.

\subsection{Ablation Study: The Oracle Gap}
The ablation (moderate regime, $n{=}6$) quantifies the oracle
(Table~\ref{tab:ablation}).
Trigger knowledge lowers ASR by $0.37$ ($p{=}0.016$) at no utility change.
\emph{Prior near-zero-ASR results for this family rely on knowing the trigger;
without it ASR is ${\sim}0.49$.}

\begin{table}[t]
\centering
\caption{RepGuard ablation (moderate regime, $n{=}6$).}
\label{tab:ablation}
\begin{tabular}{lcc}
\toprule
Variant & ASR & macro-F1 \\
\midrule
RepGuard (trigger-agnostic) & $0.490{\pm}.280$ & $0.655{\pm}.023$ \\
\textbf{Oracle canary (knows trigger)} & $\mathbf{0.120{\pm}.073}$ & $0.667{\pm}.020$ \\
$-$ remove root anchoring & $0.712{\pm}.361$ & $0.682{\pm}.012$ \\
$+$ trust memory only & $0.388{\pm}.250$ & $0.657{\pm}.020$ \\
$+$ Weibull calibration & $0.372{\pm}.343$ & $0.664{\pm}.022$ \\
\bottomrule
\end{tabular}
\end{table}

\subsection{Component Analysis}
Among trigger-agnostic components, root anchoring
is the dominant robustness lever: removing it raises ASR from $0.490$ to $0.712$,
while trust memory ($0.388$) and Weibull calibration ($0.372$) give smaller,
high-variance gains at unchanged utility (Table~\ref{tab:ablation}).

\subsection{Hyperparameter and Training-Strategy Analysis}
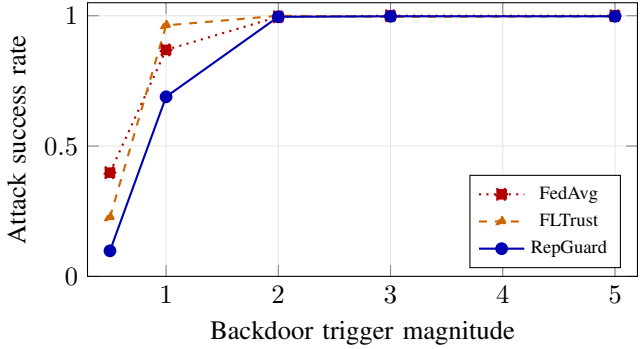
\begin{figure}[t]
\centering
\begin{tikzpicture}
\begin{axis}[width=\columnwidth,height=5.2cm,xlabel={Backdoor trigger magnitude},
ylabel={Attack success rate},ymin=0,ymax=1.05,xmin=0.3,xmax=5.2,
legend pos=south east,legend style={font=\scriptsize},grid=both,
grid style={gray!20}]
\addplot[mark=square*,red!70!black,dotted,thick] coordinates {(0.5,0.398)(1,0.869)(2,0.996)(3,1.0)(5,1.0)};
\addplot[mark=triangle*,orange!80!black,dashed,thick] coordinates {(0.5,0.226)(1,0.963)(2,1.0)(3,0.996)(5,0.998)};
\addplot[mark=*,blue!70!black,thick] coordinates {(0.5,0.098)(1,0.689)(2,0.996)(3,0.998)(5,0.998)};
\legend{FedAvg,FLTrust,RepGuard}
\end{axis}
\end{tikzpicture}
\caption{Controlled sweep ($n{=}6$): only trigger magnitude varies. Defenses help
at small triggers and collapse together at magnitude~$\geq 2$. Macro-F1 is flat
across magnitudes ($\approx 0.63$/$0.68$/$0.69$).}
\label{fig:sweep}
\end{figure}

Figure~\ref{fig:sweep} isolates trigger magnitude (all else fixed). For small
triggers RepGuard suppresses the backdoor \emph{without any oracle} (ASR $0.10$ vs
FLTrust $0.23$ vs FedAvg $0.40$ at magnitude $0.5$); by magnitude $\geq 2$ the
defended and undefended curves collapse at ASR${\approx}1.0$.

Correctly-accounted DP-SGD shows the same pattern on the privacy axis
(Table~\ref{tab:dp}): even at a strong guarantee ($\varepsilon{\approx}0.32$)
macro-F1 only falls from $0.654$ to $0.623$, but ASR stays $\geq 0.73$.
Consistent with \cite{bagdasaryan2020backdoor}, DP-SGD provides privacy at
modest utility cost but does not reduce backdoor ASR.

\begin{table}[t]
\centering
\caption{Per-example DP-SGD on CSE-CIC-IDS2018 ($\delta{\approx}10^{-5}$, $n{=}6$).}
\label{tab:dp}
\begin{tabular}{cccc}
\toprule
$\sigma$ & $\varepsilon$ & macro-F1 & ASR \\
\midrule
0.5 & 16.5 & 0.654 & 0.90 \\
1.0 & 2.14 & 0.637 & 0.77 \\
2.0 & 0.73 & 0.653 & 0.78 \\
4.0 & 0.32 & 0.623 & 0.73 \\
\bottomrule
\end{tabular}
\end{table}

\subsection{Error Analysis}
Failures concentrate where the data give the model no signal: on
NF-UNSW-NB15-v2 the rarest classes (Worms: 5 samples) are effectively
unlearnable for every method, and undefended FedAvg's low ASR on NF-ToN-IoT
reflects a weak classifier rather than robustness. Seed variance is high
($\pm0.10$--$0.31$) on the stress sweeps and the $\alpha$ trend is
non-monotonic, so single-seed readings of any cell in Table~\ref{tab:stress}
would be unreliable.

\subsection{Statistical Significance and Reliability}
All headline claims rest on $n{=}6$ paired tests; marginal results
($p{\approx}0.06$--$0.08$) are reported as trends, and the RepGuard+ ASR edge on
NF-UNSW-NB15-v2 is explicitly reported as \emph{not} significant ($p{=}0.20$).
Expanding from an initial 3-seed pilot ($n{=}3$) to $n{=}6$ seeds revised the
$\alpha{=}0.1$ gap (RepGuard ASR $0.36 \to 0.48$), and an initial $3$-seed pilot
had suggested a much larger, significant gap on NF-UNSW-NB15-v2 that did not
survive $6$ seeds; I therefore require $n{=}6$ throughout. This is direct
evidence that multi-seed evaluation is critical for FL defense benchmarks.

\section{Discussion}

\subsection{Interpretation of Findings}
Trust-based defenses preserve utility under non-IID poisoning and \emph{raise the
cost} of a backdoor (forcing larger, more detectable triggers) but do not prevent
one without trigger knowledge. The effect is
dataset-dependent: on NF-UNSW-NB15-v2 trust-based defenses cut a strong backdoor
relative to no defense (FedAvg $0.999$ vs $0.40$--$0.59$, \S\ref{sec:unsw}), with
RepGuard+ lowest, whereas on CSE-CIC-IDS2018 the same trigger defeats all methods.

\subsection{Comparison with Previous Studies}
The findings are consistent with known FL-backdoor vulnerability
results \cite{xie2023vulnerability,bagdasaryan2020backdoor}: static defenses help
only for small triggers, and DP curbs backdoors only at a utility cost. The
failed P4P transfer (\S\ref{sec:p4p}) further shows that published FL-IDS
defense numbers are hard to reproduce without released code.

\subsection{Implications}
Evaluations of FL-IDS defenses should report trigger-magnitude sweeps,
trigger-agnostic ASR, accounted $(\varepsilon,\delta)$, and multi-seed paired
tests as a minimum; the released protocol makes this the default rather than the
exception.

\subsection{Strengths}
A leakage-free, oracle-free, statistically-tested protocol; a defense-aware
adaptive attacker; three datasets spanning balanced and extreme-imbalance
regimes; and transparent reporting of negative and non-significant results.

\subsection{Limitations}
A single
MLP architecture and one (additive) trigger family; trust defenses assume a clean
disjoint root and a known bound on $f$; baselines span classic robust
aggregation, and the best-effort reimplementation of the 2026 P4P defense
\cite{p4p2026} did not transfer (\S\ref{sec:p4p}), so no comparison is drawn from
it; a faithful comparison awaits the authors' code. Other recent defenses
\cite{flidspoison2026,nswa2026,weidetect2025} are discussed but not re-implemented;
and the adaptive attacker, though defense-aware and $\lambda$-tuned, is weaker than
recent dynamic/triggerless attacks \cite{triggerless2026}.

\subsection{Future Directions}
Extending the protocol to deeper architectures, additional trigger families, and
dynamic/triggerless attacks; and re-running the comparison once P4P's official
code is available.

\section{Conclusion}
Under a leakage-free, oracle-free, statistically-tested protocol, trust-based
FL-IDS defenses are best understood as utility-preserving aggregators that help
against moderate backdoors and fail against strong ones. I quantify how much of
prior reported robustness depended on trigger knowledge, and release the protocol
for honest comparison.

\bibliographystyle{IEEEtran}
\bibliography{references}
\end{document}